
%
%

\input harvmac.tex
\input psbox.tex
\psfordvips
\def\dirac{\partial \!\!\! /}
\def\LR{$SU(2)_L \otimes
SU(2)_R \otimes U(1)_{B-L}$}
\def\SM{$SU(2)_L \otimes U(1)_Y$}

\def\PR { {\sl Phys. Rev.} }
\def\NP { {\sl Nucl. Phys.} }
\def\PL { {\sl Phys. Lett.} }

\Title{\vbox{\baselineskip12pt\hbox{ULB-TH-08/93}
\hbox{July 93}}}
{\vbox{\centerline{Spontaneous Baryogenesis with}
	\vskip6pt\centerline{Observable CP Violation}}}

\bigskip
\centerline{L. Reina\footnote{$^\natural$}{Chercheur IISN, e-mail:
lreina@ulb.ac.be} and
M. Tytgat\footnote{$^\flat$}{Aspirant FNRS, e-mail: mtytgat@ulb.ac.be}}
\bigskip
\centerline{Service de Physique Th\'eorique}
\centerline   {Universit\'e Libre de Bruxelles, CP 225}
\centerline{Boulevard du Triomphe, B-1050 Bruxelles, Belgium}
\vskip 1in
Using a left-right symmetric model with Spontaneous CP Violation and
the hypothesis of a weakly first order electroweak phase transition we
derive a relation between the produced baryon asymmetry and the
observed parameter $\varepsilon$, describing CP violation in the $K$
system.  \vskip .3in
\vfill
\break

\newsec{Introduction.}

The Electroweak Baryogenesis models came as a response to the
discovery of the existence in the Early Universe of rapid baryon
number violating anomalous processes\ref\manton{F.Klinkhamer and N.
Manton, \PR {\bf D30}, 2212 (1984)}, which could have erased any
pre-existing asymmetry. In this way the problem of preserving some
post-inflationary baryon asymmetry until now \ref\fuku{M. Fukugita and
T. Yanagida, \PR {\bf D42} (1990) 1285; J. Harvey and M.Turner,
\PR{\bf D42} (1990) 3344; B. Campbell, S. Davidson, J.Ellis and K.
Olive, \PL{\bf B256} (1992) 457} can be replaced by the problem of
generating the baryon asymmetry of the universe (BAU) at the latest
possible stage: the electroweak phase transition (EWPT) \ref\all{V.
Kuzmin, V.  Rubakov and M. Shaposhnikov, \PL{\bf B155}, 36 (1985)}.
This scheme has also opened the exciting possibility to further test
the standard model (SM) and probe its numerous extensions.

The task of baryogenesis is twofold: first one has to insure that
Sakharov's conditions \ref\sakh{A.D. Sakharov, JETP Lett. {\bf 6},24
(1967) } are fulfilled and then check that an efficient mechanism is
realised. It is very nice and intriguing that the SM can satisfy by
its own the former point: C and CP violation, B non-conservation and
an out-of-equilibrium stage through a first order phase transition
{\all}. Whether this model is able to generate the right asymmetry is
still an open question \ref\shapo2{G.R. Farrar and M.E. Shaposhnikov,
CERN preprint TH.6732 (1993)} and the most popular point of view is to
consider the baryon excess of the universe as an hint for something
standing beyond the SM (see
\ref\andiii{A. G. Cohen, D.B. Kaplan and A.E. Nelson, UCSD-PTH-93-02 preprint,
(1993)} and \ref\turok{N. Turok,Imperial/TP/91-92/33 preprint (1992)} and
references therein).

In this letter we investigate the left-right model (LR) based on the
gauge group $SU(2)_L
\otimes SU(2)_R \otimes U(1)_{B-L}$, already considered in
\ref\moh{R.N. Mohapatra and
X. Zhang, \PR {\bf D46}, 5331 (1992)} and \ref\moi{J.-M. Fr\`ere, L.
Houart, J.M.  Moreno, J. Orloff and M. Tytgat, CERN preprint TH.6747
(1992)}. One of the most interesting suggestions is that, if CP is
spontaneously violated at the electroweak symmetry breaking scale,
then it should be possible to relate the baryon asymmetry of the
universe (BAU) to the low energy CP violation phenomenology and in
particular to the precisely measured $\varepsilon$ parameter in the
$K^0$ system.

We think this model deserves further attention since it offers a
common source to the, until now, only two {\it well measured}
manifestations of CP violation in nature.

In {\moi} it was supposed that the electroweak phase transition was
strongly first order occurring through nucleation and expansion of
bubbles of true vacuum with {\bf thin} walls. The reflection of quarks
on the moving wall in a CP violating way creates a charge excess which
is converted into a baryon excess by the rapid B violating processes
occurring in the symmetric phase\foot{This is the so called Charge
Transport Mechanism proposed in \ref\andy1{A.G. Cohen, D.B. Kaplan and
A.E. Nelson, \NP {\bf B373}, 453 (1992)}}.

Here we will follow the opposite hypothesis, {\it i.e.} that the
expanding bubbles have a {\bf thick} wall. The appropriate mechanism
here is the so called Spontaneous Baryogenesis
\ref\andi2{A.G. Cohen, D.B. Kaplan and A.E. Nelson, \PL {\bf 245B},
561 (1990)} in which
the BAU is created within the wall. We will get a strong upper bound
for the produced baryon asymmetry in terms of the $\varepsilon$
parameter and purely dynamical quantities ($\kappa$, $m_{top}$, $M_2
\approx M_R$ and the signature of the quark mass matrices). If we
further suppose that this LR model `` saturates" $\varepsilon$, we get
the nice result that the BAU and Re $\varepsilon$ must have the same
sign.

\newsec{Spontaneous CP violation in the LR model}

The matter content of the model and the symmetry properties imposed on
the Lagrangian are given in the Appendix. The salient points are:

\item{-} the explicit LR symmetry is broken to the SM one at a scale
$M_R$ = O(TeV)
through the $vev$ of a scalar field transforming as a triplet under
$SU(2)_R$;
\item{-} the electroweak symmetry is broken through the $vev$ of a
scalar bi-doublet
$i.e.$ a field in the $(1/2, 1/2, 0)$ representation of the gauge
group.

\noindent The Yukawa couplings of quarks read:
\eqn\yuk{{\cal L}_{\hbox{{Yukawa}}} = - \Gamma_{ij}\, \bar q_{iL}\, \phi
\, q_{jR} - \Delta_{ij}\, \bar q_{iL}\tilde \phi\, q_{jR}\quad , }
where $\phi$ is the bi-doublet field, with $vev$:
\eqn\bida{\langle\phi\rangle = \left(\matrix{v & 0 \cr 0 & w \cr}\right)}
and $\tilde\phi \equiv \sigma_2 \phi^*\sigma_2$. The coupling matrices
$\Delta$ and $\Gamma$ are real and symmetric due to the assumed CP
invariance of the original Lagrangian.

Spontaneous CP violation occurs if $v$ and $w^*$ are relatively
complex.  After a $SU(2)_L$ or $SU(2)_R$ transformation it is possible
to express this by one single phase $\alpha$:
\eqn\bidb{\langle\phi\rangle = e^{i\,\alpha/2}\left(\matrix{|v| & 0
\cr 0 & |w| \cr}\right)}
\noindent As noted in \ref\branco{G.C. Branco and L. Lavoura, \PL {\bf
165B}, 327 (1985)}
the simplest model cannot give rise to a non trivial phase without
fine tuning, but the minimal extension of adding a pseudo-scalar
singlet suffices to solve the problem. This is briefly summarised in
the Appendix.

In the chosen phase convention the mass matrices of the quarks are
given by
\eqn\mass{\eqalign{M^{(u)} &= v \Gamma + w^* \Delta = |v|
e^{i\,\alpha/2}
(\Gamma + r e^{-i\,\alpha} \Delta)\cr M^{(d)} &= w \Gamma + v^* \Delta = |v|
e^{i\,\alpha/2}(r\Gamma + e^{-i\,\alpha} \Delta)\cr}} with $w/v^* =
|w/v|\, e^{i\,\alpha}= r\,e^{i\,\alpha} $. These are complex symmetric
matrices which can be diagonalized by two unitary matrices:
\eqn\rot{\eqalign{M^{(u)} &= e^{i\,\alpha/2} U D^{(u)} U^T \cr
M^{(d)} &= e^{i\,\alpha/2} V D^{(d)} V^T \cr}} At the difference of
the SM the L and R quarks are not rotated independently. This has as
important consequence that the mixing matrices $K_L$ and $K_R$ (the
generalisation of the KM matrix) are not independent and in the basis
(phase convention) we have chosen they are related by \eqn\KM{K_L =
U^{\dag}V = K_R^*} Note that the factorisation of $e^{i\,\alpha/2}$ in
{\rot} leaves {\KM} invariant. In the same way one can factorize
$e.g.$ $|v|$ without changing the rotation matrices $U$ and $V$. Note
also that, for the case of $N_f$ flavours, they are $(N_f^2 - N_f +1)$
physical phases in the mixing matrices and so it is possible to have
CP violation already with two generations. The nicety of the LR model
with spontaneous CP violation is that all the phases are functions of
$\alpha$ and that in the limit of small $r \sin \alpha$ they are
analytically calculable.  Furthermore, an enhancement of the
$\Delta$S$=2$ channel ensures a small value of
$\varepsilon^\prime/\varepsilon$ for dynamical reasons
\ref\JMJM{G. Branco, J.-M. Fr\`ere and J.-M. Gerard, \NP {\bf B221},
317 (1983)}.

The key point, as remarked by Chang \ref\chang{D. Chang, \NP {\bf
B214}, 435 (1983)}, is that in the limit where $r\sin\alpha$ goes to
zero in {\mass} there is no CP violation. With the hypothesis of small
$y = r\sin\alpha$ ($a\,posteriori$ verified) one may expand $K_L$ as
follows
\eqn\dev{K_L = e^{-i\,\alpha/2}(K_0 + i y K_1) + O(y^2)}
where the lowest order matrix $K_0$ is real and experimentally known
being equal to $\vert K_{KM}\vert$, up to signs. As shown in {\chang}
$K_1$ is calculable as function of the measured mixing angles and
quark masses.

We will now show  that it is possible to use the vacuum (at T=0)
results to describe the features of CP violation during the EWPT, provided:
\item{-}$r$ is constant during the phase transition;
\item{-}$\alpha$ is small.

\noindent Let us define from {\mass} the following matrices:
\eqn\masst{\eqalign{\tilde M^{(u)} &= \Gamma + r e^{-i\,\alpha} \Delta
=\Gamma +r\cos
\alpha \Delta - i r \sin \alpha \Delta = U \tilde D^{(u)}U^T\cr
                   \tilde M^{(d)} &= r\Gamma + e^{-i\,\alpha} \Delta
=e^{-i\alpha}(r\cos \alpha\Gamma + \Delta + i r \sin \alpha
\Gamma)=V\tilde D^{(d)}V^T\cr}} As the $\tilde D$'s are dimensionless
they can only be function of $r$ and $\alpha$. Chang's expansion of
$U$, $V$, $D^{(d)}$ and $D^{(u)}$ is given by
\eqn\matexp{\eqalign{U &= U_0 - i y U_1 + O(y^2) \cr
                     V &= e^{-i\, \alpha/2}(V_0 + i y V_1) + O(y^2)
\cr D^{(u,d)} &= D_0^{(u,d)} + i y D_1^{(u,d)} + O(y^2)\cr}} It can be
shown {\chang}, \ref\ecker{G. Ecker and W. Grimus, \NP {\bf B258}, 328
(1985)} that these matrices have the following properties:
\item{-}$U_0$ and $V_0$ are orthogonal, they diagonalize the real part
of {\masst} and
$K_0 = U_0^\dagger V_0$; \item{-}$D_1^{(u,d)}=0$;
\item{-}$U_0^\dagger U_1$ and $V_0^\dagger V_1$ are real and symmetric.

\noindent As appears from {\masst} and {\matexp} if $r$ is constant during
the EWPT and $\alpha$ is small, $U_{0}$, $V_0$ and $\tilde D_0^{(u,d)}$ are
constant to order $\alpha$. Moreover one has from {\ecker}:
\eqn\eck
{\eqalign{(1-w_{\alpha}^2)(U_0^\dagger U_1)Q{ij} & = {(K_0
D^{(d)}K_0^T)_{ij}\over m_i^{(u)}+ m_j^{(u)}} - 1/2\, w_\alpha
\delta_{ij}\approx{(K_0 \tilde D_0^{(d)}K_0^T)_{ij}\over \tilde m_i^{(u)}+
\tilde m_j^{(u)}} - 1/2\,
 r  \delta_{ij}\cr  (1-w_{\alpha}^2)(V_0^\dagger V_1)Q{ij} &
= {(K_0^T D^{(u)}K_0)_{ij}\over m_i^{(d)}+ m_j^{(d)}} - 1/2\, w_\alpha
\delta_{ij}\approx{(K_0^T \tilde D_0^{(u)}K_0)_{ij}\over \tilde m_i^{(d)}+
\tilde m_j^{(d)}} - 1/2\, r \delta_{ij} \cr}}
where $w_\alpha = r
\cos\alpha \approx r$ to the same order.
One concludes that $U_0^\dagger U_1$ and $V_0^\dagger V_1$ are also
almost constant during the EWPT. Hence the matrices $U_{(0,1)}$,
$V_{(0,1)}$ and $\tilde D^{(u,d)}$ introduced above are the same as in
the $T=0$ vacuum. All the effects of CP violation during the EWPT can
be parametrized by the sole variation of the phase $\alpha$ or more
properly of the small expansion parameter $y$. This property will be
extensively used below but we want to show first that our hypotheses
of constant $r$ and small $\alpha$ may indeed be easily satisfied.

The $vev$ of $\alpha$ and $r$ are given by the minimisation of the
most general scalar potential compatible with LR and CP symmetry (see
the Appendix) minimally extended to give rise to spontaneous CP
violation {\branco}:
\eqn\potalpha{\tan \alpha = {C_1 \eta^2 \over C_2 \eta^2 + B_4\,\sigma_R^2}}
and with $\tan s \equiv 1/r$,
\eqn\potr{\tan 2s = {(B_4 \cos\alpha)\, \sigma_R^2 + (C_1 \sin \alpha
+ C_2 \cos \alpha)\,
\eta^2 \over A_3 \sigma_R^2}}
where $\eta$ and $\sigma_R$ are the $vev$'s of the pseudoscalar
singlet and the right scalar triplet respectively. The other
parameters are combinations of dimensionless couplings in the scalar
potential. For $\eta =O(100$ GeV), corresponding to the requirement of
spontaneous CP violation at the EW scale and $\sigma_R = O$(1 TeV), a
generic value for the R scale,
\eqn\alp{\tan\alpha \approx C_1 \eta^2 / B_4 \sigma_R^2}
has a well defined sign and is monotonically varying from $0$ to some
{\bf small} finite value as $\eta$ varies from $0$ to its $vev$ during
the EWPT. Also, as $\alpha$ is small, \eqn\rrr{\tan 2s \approx B_4 /
A_3 = \hbox{constant}.} The temperature dependent corrections to the
above results can only be logarithmic and should consequently barely
change the conclusions.

\newsec{Spontaneous Baryogenesis with Spontaneous CP Violation}
In {\moi} it was argued that the phase transition in the model we consider
could be strongly first order because of the presence of trilinear couplings in
the scalar potential (cf Appendix). However, the astonishing
complexity of the potential cannot allow us to exclude the opposite
possibility, $i.e.$ that, for some values of the parameters, the EWPT
is weakly first order, occurring through nucleation and expansion of
bubbles of true vacuum with a thick wall.

The mechanism in {\moi} used the reflection and transmission of heavy
quarks on the wall in a CP violating way. If the wall is thick, the
reflection probability at threshold is exponentially suppressed and
the mechanism breaks down. A possible alternative is to create the
baryon asymmetry within the wall. For this we will suppose that all
the processes, but baryon number violating ones, are in quasi thermal
equilibrium within the wall. This means that the thermalization time
of the particles $\tau_{th}\approx (0.25-0.08\, T)^{-1}$ {\andiii} is
significantly smaller than the characteristic time of the moving wall
$\tau_{wall} = \delta_w / v_w$, where $\delta_w$ and $v_w$ are
respectively the thickness and the velocity of the wall. The
characteristic time for baryon number violation $\tau_B$ is estimated
to be of the order $(\alpha_W^4 T)^{-1}\approx 10^6/T$ in the
symmetric phase and is much longer in the broken phase. Baryon number
violating processes are thus always out of thermal equilibrium. We
further suppose that within the wall $\ddot\alpha \ll \dot \alpha$ so
that the phase is quasi-static.

The hypothesis $\tau_{th} \ll \tau_w$ with $\ddot\alpha\ll\dot\alpha$
is referred as the {\it adiabatic regime} in \andiii. Thermal
equilibrium is maintained through the bubble wall by fast
interactions, while chemical equilibrium may not be satisfied as some
processes, among which baryon number violation, are comparatively
slower. The departure from chemical equilibrium is handled by
introducing effective chemical potentials due to the interaction with
the slowly varying phase $\alpha$.

To see how this may happen we go back to the Yukawa Lagrangian for the
quarks
\yuk. In the rest frame of the plasma, the time varying phase of the bi-doublet
scalar field may be removed by a time dependent rotation of the quarks
fields.  We choose the rotation which diagonalizes the couplings with
the scalar field
\eqn\roti{\eqalign{u_L \longrightarrow u_L^\prime &=
e^{-i\,\alpha/4}U^\dagger u_L\cr
                  u_R \longrightarrow u_R^\prime &= e^{i\,\alpha/4}
U^T u_R\cr d_L \longrightarrow d_L^\prime &= e^{-i\,\alpha/4}
V^\dagger d_L\cr d_R \longrightarrow d_R^\prime &= e^{i\,\alpha/4} V^T
d_R\cr}} where we have used the convention of {\rot} and {\KM}. The
first consequence of {\roti} is that the kinetic part of the quark
Lagrangian gives rise to a new term: \eqn\kine{{\cal L}_K
\longrightarrow {\cal L}^\prime_K\, + \,{\cal L}_{eff}} where
\eqn\eff{\eqalign{{\cal L}_{eff} = &-{1 \over 4}\left \{ \bar
d^\prime_L (\dirac\alpha)
d^\prime_L \,- \,\bar d^\prime_R (\dirac\alpha) d^\prime_R \, + \,
\bar u^\prime_L (\dirac\alpha) u^\prime_L \, -\,\bar u^\prime_R
(\dirac\alpha) u^\prime_R\right \}\cr &+ \,i\,\bar
d^\prime_L(V^\dagger\dirac V)d^\prime_L \,+ \,i\,\bar
d^\prime_R(V^T\dirac V^*)d^\prime_R \cr &+
\,i\,\bar u^\prime_L(U^\dagger\dirac U)u^\prime_L \,+ \,i\,\bar
u^\prime_R(U^T\dirac U^*)u^\prime_R}} Now we use the expansions
{\matexp} of $U$ and $V$ to get:
\eqn\effii{\eqalign{{\cal L}_{eff} = &-{1 \over 4}\left \{ -\bar
d^\prime_L (\dirac\alpha)
d^\prime_L \,+ \,\bar d^\prime_R (\dirac\alpha) d^\prime_R \, + \,
\bar u^\prime_L (\dirac\alpha) u^\prime_L \, -\,\bar u^\prime_R
(\dirac\alpha) u^\prime_R\right \}\cr &-\,\bar d^\prime_L (V_0^\dagger
V_1\dirac y) d^\prime_L\, + \,
\bar d^\prime_R (V_0^\dagger V_1\dirac y)^* d^\prime_R \cr
& -\,
\bar u^\prime_L (U_0^\dagger U_1\dirac y) u^\prime_L \, +\,
\bar u^\prime_R (U_0^\dagger U_1\dirac y)^* u^\prime_R\cr}}
where we have used that $U_{(0,1)}$ and $V_{(0,1)}$ are constant
matrices. As $y$ is time dependent, this new part of the Lagrangian is
of the form
\eqn\potchem{{\cal L}_{eff} = - \sum_i \mu_i\, \bar q_i\gamma_0 q_i }
where the summation is over flavours and chiralities, and $\mu$ is
proportional to $\dot y $: the $\mu$'s look like chemical potentials
{\andiii} though being dynamical quantities rather than Lagrange
multipliers as in thermodynamics.

The quark distributions will try to minimize the effective potential,
possibly through baryon number violating processes, that could lead to
a net baryon excess at the end of the EWPT {\andiii}. However we now
show why this does not happen here. First, rapid chirality changes
occur both by emission and absorption of real scalar particles and by
interaction with the chirality non-conserving external scalar field.
Consequently baryon violating processes act equally on left-handed and
right-handed quarks. Then the baryon creation rate is
proportional to the rate for B violation times the sum of the
effective chemical potentials or, in the above formulation, to the
trace of {\it all} the matrices in {\effii} : \eqn\naivB{\eqalign{\dot
n_{B} \propto \Gamma_B\,
\biggl\{   \left(3/4\dot\alpha - \dot y Tr(V_0^\dagger V_1)\right)_{d_L} &+
\left(-3/4\dot\alpha + \dot y Tr(V_0^\dagger V_1)^*\right)_{d_R}\cr
\left(-3/4\dot\alpha + \dot y Tr(U_0^\dagger
U_1)\right)_{u_L} &+ \left(3/4\dot\alpha - \dot y Tr(U_0^\dagger
U_1)^*\right)_{u_R} \biggr\}\cr}} Finally, as $U_0^\dagger U_1$ and
$V_0^\dagger V_1$ are real matrices, the different contributions
cancel to zero. This can be understood by the fact that parity
violation is one of the ingredient required for baryogenesis {\sakh}:
the cancellation between the left-handed and the right-handed terms is
then a remnant of the P-conserving structure of our LR model. The
following figure gives an equivalent (naive) illustration of this
phenomenon: $${\psboxscaled{650}{LRfigure.1}} $$

The previous conclusions would be disastrous if not for the presence
of a second contribution. Indeed our rotation, being chiral, has a
gauge anomaly. The Lagrangian is then modified as
\eqn\anomaly{{\cal L} \longrightarrow {\cal L}+{\cal L}_\Theta\,={\cal
L}\,+\,{\Theta (x) \over 16 \pi^2} TrW_{\mu\nu}\tilde W^{\mu\nu}}
where \eqn\thet{\Theta (x)	= \hbox{Arg det} M^{(u)}M^{(d)}} and
$W_{\mu\nu} = g / 2i \sum_a \lambda_a W_{\mu\nu}^a $, $\lambda_a/2$
being the gauge group generators in the appropriate representation. A
priori such a term arises for each gauge group of the model, but the
only ones relevant for baryogenesis are $SU(2)_L$ and $SU(2)_R$
because they are chiral and have a non-trivial topological structure.
Moreover, as the $SU(2)_R$ symmetry is broken, there is a high energy
barrier (a R-sphaleron {\moi}) between the different R-vacua and
topological fluctuations are exponentially suppressed in this sector
of the model at the electroweak temperature. So, as in the SM,
``rapid" baryon number violation will occur through configurations of
$SU(2)_L$.

Note that {\anomaly} also gives a contribution in the strong sector.
Because parity is spontaneously broken, the LR model belongs to a
class of models in which the $\Theta_{strong}$ is {\it a priori}
calculable, but the final value of $\Theta$ we will obtain largely
exceeds the bound from the $e.d.m.$ of the neutron
\ref\marciano{S.M. Barr and W. J. Marciano, in CP Violation, C.
Jarlskog ed., World
Scientific, 455 (1988)}. Some fine tuning -- $i.e.$ the introduction
of a non zero $\Theta_{strong}$ bare -- is thus necessary in order to
cancel the $\Theta_{strong}$. As our mechanism of baryogenesis is only
sensitive to the {\it variation} of $\Theta$, this will not affect our
final result.

{}From the anomaly equation
\eqn\ano{\partial_{\mu} j^\mu_5\, = -\,{1\over 8 \pi^2}\, Tr W^L_{\mu\nu}\tilde
W_L^{\mu\nu}} we deduce the divergence of the baryonic current:
\eqn\barcur{\partial_{\mu} j^\mu_B\, =-{N_f \over
2}\partial_{\mu} j^\mu_5\, = {N_f\over 16 \pi^2}\, Tr
W^L_{\mu\nu}\tilde W_L^{\mu\nu}} where $N_f$ is the number of
flavours\foot{Notice that if we take into account the anomalous
contribution of the right gauge bosons, then {\barcur} becomes
\eqn\barcuri{\partial_{\mu} j^\mu_B\, = {N_f\over 16 \pi^2}\lbrace
Tr W^L_{\mu\nu}\tilde
W_L^{\mu\nu} - Tr W^R_{\mu\nu}\tilde W_R^{\mu\nu} \rbrace.} As the
$\Theta$-term acts the same for the L and R sector their contributions
would cancel as is the case for the effective chemical potentials. The
breaking of the LR symmetry at an higher scale is really crucial here
in order to satisfy Sakharov's conditions.}. Using the anomaly
equation with {\anomaly} we may get an effective potential for the
baryon number density
\eqn\potB{{\cal L}_\Theta = {\Theta \over N_f}\, \partial_\mu j_B^\nu = -\,
{\dot\Theta\over N_f}\, j^0_B.} The fermions will try to minimize the
potential {\potB} through baryon number violating processes. This, as
pointed in \andiii, is equivalent to the introduction of chemical
potentials constraining the processes in a system slightly
out-of-chemical-equilibrium.

The master equation for the baryon number density is then given by
\ref\willy{M. Dine, O. Lechtenfeld, B. Sakita, W. Fishler and J.
Polchinsky
\NP {\bf B342}, 381 (1990)} \eqn\botlz{{d \,n_B \over d\, t}\, =\, - 3 N_f
{\Gamma_B \over T}{\partial F \over \partial B}} where $F$ is the free
energy of the system.  If $\Theta$ is quasi static during the EWPT,
then ${\partial F /
\partial B}\approx {\partial V / \partial j^0_B}= \dot\Theta /N_f$.

As we do not know the details of the phase transition, hence the exact
time dependence of $\Theta$, and as little is known on the exact rate
for B violation in the intermediate regime of the phase transition, we
will make the following simplifications:
\item{-}the rate per unit volume will be taken to be constant
and the same as in the symmetric phase, $\Gamma_B = \kappa
\,\alpha_W^4 T^4$ were $\kappa$ parametrizes our ignorance of the
exact rate but has been estimated from numerical simulations to be
O(1);
\item{-}the variation of $\Theta$ during the EWPT
takes its maximal value.

\noindent These assumptions clearly give a very conservative upper
bound for the
baryon number produced during the phase transition. Using the
expression {\rot} for the mass matrices and the expansion {\matexp}
for the rotation matrices, we have the following expression for
$\Theta (x)$:
\eqn\thetaexp{\eqalign{\Theta (x) &= \hbox{Arg det}
M^{(u)}M^{(d)} = \hbox{ImTr ln}M^{(u)}M^{(d)}\cr &= \hbox{ImTr
ln}\lbrace 1 - 2\, i\, y(x) (U_0^\dagger U_1 - V_0^\dagger V_1)\rbrace
+ O(y^2)\cr &= 2\, y(x)\, Tr(V_0^\dagger V_1 - U_0^\dagger U_1)\cr}}
where we have used the small $y$ approximation and the properties of
the matrices given in the previous section. Note also that
$U_0^\dagger U_1$ and $V_0^\dagger V_1$ are explicitly given by {\eck}
and are {\it a priori} experimentally accessible quantities.

With the above assumptions we get the following upper bound on the
baryon-to-entropy ratio:
\eqn\BAU{\eqalign{{n_B \over s} &\leq {n_B \over s}\bigr\vert_{max} =
- 3\,\, {\kappa
\alpha_W^4 T^3 \over s}\, (\Theta_f - \Theta_i)\cr  &\leq - \kappa \,
{135 \over g_{*S}
\pi^2}\, \alpha_W^4\, r \sin \alpha\, Tr(V_0^\dagger V_1 - U_0^\dagger
U_1)\cr}}
where we have used the fact that $\sin \alpha$ starts from zero (see
{\alp}) and $s = 2
\pi^2 /45 g_{*S} T^3$ is the entropy density. Note that {\BAU} does
not depend of the bare value of $\Theta$.

\newsec{Phenomenological Implications:
{\bf $\varepsilon$} and the BAU} The $\varepsilon$ parameter as
predicted in the LR model with spontaneous CP violation is given by
{\ecker}, \ref\JM{J.-M. Fr\`ere, J. Galand, A. Le Yaouanc, L. Olivier,
O.  P\`ene and J.-C. Raynal \PR {\bf D46}, 337 (1992)}:
\eqn\epsi{\varepsilon =
\varepsilon_{SM} + \varepsilon_{LR}} where
\eqn\epssm{\varepsilon_{SM} = e^{i\,(\pi/4)}\, 1.34\, s_2\, s_3 \sin
\delta \left\{ 1 +
860\, S\left ( {m_t^2\over M_1^2}\right)s_2\, \hbox{Re}V_{ts}\right\}}
and
\eqn\epslr{\varepsilon_{LR} = -
e^{i\,(\pi/4)}\,0.36\sin(\delta_2-\delta_1)\left [
{M_2\over \hbox{1.4 TeV}}\right ]^2\left\{1 + 0.05
\ln\left[{M_2\over\hbox{1.4 TeV}}\right]\right\}.} Some remarks are in
order here:
\item{-}the expression for $\varepsilon_{SM}$ is the same as the one
calculated in the
standard model from the KM matrix with three generations. $M_1$ is
essentially the $W_L$ boson of the SM. The product $s_2\, s_3 \sin
\delta$ is calculable in the LR model and is linear in $y =
r\sin\alpha$ {\ecker};
\item{-}In the LR model, there is  CP violation
in the $K_0-\bar K_0$ system already with two generations. In that
case, there are three CP violating phases ($\gamma$, $\delta_1$ and
$\delta_2$) in the mixing matrices which are also calculable functions
of $r$ and $\alpha$. In the case of three generations, the LR
contribution of the third generation is usually negligible in
comparison to the one of the first two families. Moreover the LR
contribution with the first two generations usually dominates the SM
one for not too heavy $W_R
\approx W_2$ {\JM};
\item{-} There is a subtlety in the fact that in a LR model the
sign of the quark masses are observable. One can remove them from the
mass matrices by doing a rotation on the right-handed quarks to the
cost of effects in the $K_R$ mixing matrix. So there is actually a
discrete set of $2^{2 N_f -1}$ different models. This clearly weakens
the predictability of the model {\JM}: for example for some signatures
there are cancellations in $\varepsilon_{LR}$ and $\varepsilon_{SM}$
so that the third generation dominates in {\epsi}\JM.

\noindent The interesting point for us is that $\varepsilon$ is
linear in $y = r \sin\alpha$, just as is our upper bound for the BAU.
For the reasons given above, a complete analysis of the consequences
for the LR model requires some care. Nevertheless, if we suppose that
the LR model with two generations ``saturates" $\varepsilon$ in the
sense that $\varepsilon_{LR}$ dominates $\varepsilon_{SM}$, we have
the following expressions:
\eqn\domin{\eqalign{\delta_2 - \delta_1 =& {r \sin\alpha \over
1 - w_\alpha^2}\biggl [ {m_u c^2 + m_c s^2 \over m_d} - {m_u s^2 + m_c
c^2 \over m_s} \cr &+ 2{ m_d - m_s \over m_u + m_c} + 2 {m_u - m_c
\over m_d + m_s} (s^2 - c^2)\biggr ] \cr
\approx & {r \sin\alpha \over 1 - w_\alpha^2}
\biggl [ 7\, sign\left ({m_c \over m_s}\right
) + 6 \,sign \left ({m_c \over m_d}\right) \biggr ]\cr}} and
\eqn\domini{\eqalign{r \sin\alpha\, Tr(V_0^\dagger V_1 - U_0^\dagger
U_1) =& {1 \over 2}{r
\sin\alpha \over 1 - w_\alpha^2}\biggl [c^2 \left ({m_u \over m_d} - {m_d \over
m_u}\right ) + c^2 \left ({m_c \over m_s} - {m_s \over m_c}\right )\cr
&+ s^2
\left ({m_c \over m_d} - {m_d \over
m_c}\right ) + s^2 \left ({m_u \over m_s} - {m_s \over m_u}\right )
\biggr ]\cr
\approx & {1 \over 2}{r \sin\alpha
\over 1 - w_\alpha^2} \biggl [ 7\, sign\left ({m_c \over m_s}\right
) + 6 \,sign \left ({m_c \over m_d}\right) \biggr ]\cr}} where $s
\equiv \sin \theta_C$ is the Cabibbo mixing angle.

We finally get from {\BAU} and {\epslr}:

\eqn\fnal{{n_B \over s} \leq 135 \, {\sqrt{2}\over 0.72}\,
{\kappa \alpha_W^4 \over \pi^2 g_{*S}}\, \hbox{Re }
\varepsilon \, \left ({1.4\, \hbox{TeV} \over M_2 }\right )^2 \left [1
-0.05 \ln \left({
M_2 \over 1.4 \,\hbox{TeV}}\right ) \right ]}

\noindent There are only two unknowns in the upper bound $i.e.$
$\kappa$, describing
the amount of B violation and $M_2$, the mass of the charged
right-handed $W$: these are dynamical parameters independent of CP
violation. $\hbox{Re } \varepsilon$ is observable and the expression
{\fnal} predicts that it must have the same sign as the BAU. {}From the
experimental value of $\hbox{Re } \varepsilon$ we get also an upper
bound on the mass $ M_2$ \ref\kolb{E. Kolb and M. Turner, {\it The
Early Universe}, Addison--Wesley, New York (1990)}: \eqn\num{4\,
10^{-11}\leq {n_B \over s} < 0.7\,\kappa \left ({1.4\,\hbox{TeV} \over
M_2 }\right )^2 10^{-9}} or
\eqn\bound{M_2 < 6\, \sqrt{\kappa}\, \hbox{TeV}}
where $\kappa$ is usually estimated to be O(1). This upper bound is
consistent with the one needed to satisfy $\varepsilon \approx
\varepsilon_{LR}$ which is $M_2 \leq 19$ TeV
\JM.

\newsec{Conclusions}

The LR symmetry coupled to spontaneous CP violation offers the
interesting opportunity to describe in a unified framework the
generation of the baryon asymmetry and the LR phenomenology of the
$K_0-\bar K_0$ system. The existence of one unique source of CP
violation naturally permits to relate $n_B/s$ and $\varepsilon$ and
this in agreement with the observations.

The upper bound on the baryon asymmetry we have obtained rests on very
few assumptions.  The most important ones are that the phase
transition is weakly first order and that spontaneous CP violation
occurs at the electroweak scale. Their possible validity is hidden by
the complexity of the scalar potential, in which we do not think it is
useful to go much further at this stage. Our very conservative bound
still indicates that the mechanism of spontaneous baryogenesis may be
in difficulty if the R scale is too high ($M_2 >$ few TeV). However we
have only considered the ``natural" case where the LR model
``explains" CP violation. Whether the conclusion is the same for
others signatures is an open question. Linked to this is the agreement
in sign we have found between $n_B/s$ and $\hbox{Re }
\varepsilon$ which could possibly not be satisfied in all the cases. Clearly
this must be complemented by a conjoint analysis of the consequences
on $\varepsilon^\prime$ and possibly the electric dipole moment of the
neutron.

Also it could be interesting to return to the thin wall case in which
baryogenesis is more efficient {\andiii} to see if it is possible to
obtain a similar relation between the sign and magnitude of $B$ and
$\varepsilon$.

\newsec{Acknowledgements}
We would like to thank J.-M. Fr\`ere and Ph. Spindel for discussions
and encouragements.

\appendix {}{} The model is based on the gauge group \LR with quarks
and leptons in fundamental representations
\eqn\rep{\eqalign{Q_L &= \left (\matrix{u \cr d}\right )_L \hskip15pt ({1/2, 0,
1/3})
\hskip40pt Q_R = \left (\matrix{u \cr d}\right )_R \hskip15pt ({0, 1/2,
1/3})\cr Le_L &= \left (\matrix{\nu \cr e}\right )_L \hskip15pt ({1/2,
0, -1})
\hskip40pt Le_R = \left (\matrix{\nu \cr e}\right )_R \hskip15pt ({0,1/2,
-1})\cr}} The generalisation of the Gell-Mann-Nishijima relation is $Q
= T_{3L} + T_{3R} + {B-L \over 2}$.  The two $SU(2)$ coupling
constants $g_L$ and $g_R$ are set to be equal by imposing the discrete
symmetry L $\longleftrightarrow$ R.

To break the symmetries down to $U(1)_Q$ different Higgs mutliplets
are introduced
\ref\desh{N.G. Deshpande, J.F. Gunion, B. Kayser and F. Olness,
 \PR{\bf D44}, 837 (1991)}:

\eqn\higgs{\eqalign{\Delta_L =(1,0,
2) \hskip15pt & \hskip15pt \Delta_R =(0,1, 2)\cr
\phi=(1/2,&1 / 2,0)\cr}}
The introduction of the scalar bi-doublet imposed by the LR structure
of the model is fundamental in the discussion of spontaneous CP
violation and makes the difference with respect to models using two
doublets for baryogenesis. The breaking of {\LR} occurs in two steps:
firstly through the vacuum expectation value of the triplet $\Delta_R$
at some O(TeV) scale; secondly \SM is broken to $U(1)_Q$ at the
electroweak scale O(100 GeV) through the $vev$ of the bi-doublet
field.

 For completeness we also sketch here the results obtained in
{\branco}. Their objective is to know if spontaneous CP violation is
possible with the matter content given above. For this the following
transformation properties under P and CP are imposed on the scalar
fields:
\eqn\P{\hbox{P} : \phi_i \longrightarrow \phi_i^\dagger\hskip15pt;
\hskip15pt\Delta_L \leftrightarrow \Delta_R}
and
\eqn\CP{\hbox{CP}: \phi_i \rightarrow \phi_i^* \hskip15pt;
\hskip15pt\Delta_L \rightarrow \Delta_L^*\hskip15pt;
\hskip15pt\Delta_R \rightarrow \Delta_R^*}
where $\phi_1 \equiv \phi$ and $\phi_2 \equiv \tilde \phi=\tau_2
\phi^* \tau_2$.  Imposing P and CP to be symmetries of the scalar
potential
\eqn\scapot{V = V_{\phi} + V_\Delta + V_{\Delta\phi}} give constraints on the
couplings \branco. The $vev$ of the scalar mutltiplets are set to
\eqn\vevs{\langle\phi\rangle = e^{i\,\alpha/2}\left(\matrix{|v| & 0 \cr 0 &
|w| \cr}\right)} and
\eqn\vevss{\langle\Delta_L\rangle = \left(\matrix{0 & 0 \cr \sigma_L
e^{i \delta} &
0 \cr}\right)\hskip30pt\langle\Delta_R\rangle = \left(\matrix{0 & 0
\cr \sigma_R & 0 \cr}\right)} It is shown in {\branco} that, without
fine tuning, $\alpha$ must be $0$ or $\pi$. In order to have
spontaneous CP violation a simple extension is proposed: a
pseudoscalar singlet is introduced which transforms under P and CP as
follows:
\eqn\PCP{\hbox{P}: \eta \rightarrow -\eta \hskip50pt \hbox{CP}: \eta
\rightarrow -\eta
} This allows one to introduce in the scalar potential the following
additional terms
\eqn\poten{V_{\eta\phi} = i\, C_1 \,\langle \eta\rangle \,\eta\,
Tr(\phi_2^\dagger
\phi_1 - \phi_1^\dagger \phi_2)\,+\, C_2\, \eta^2 Tr(\phi_2^\dagger \phi_1 +
\phi_1^\dagger \phi_2)}
which contain a term odd in $\alpha$. Then a non zero $vev$ is
obtained for $\alpha$ by minimizing the potential:
\eqn\tanalpha{\tan \alpha = {C_1 \eta^2 \over C_2 \eta^2 + B_4\,\sigma_R^2}}
where $B_4$ is simply a combination of couplings from
$V_{\Delta\phi}$.  As this is used in the body of the paper we also
give the expression for the ratio $|v/w| =1/r = \tan s$:
\eqn\tandeuxs{\tan 2s = {(B_4 \cos\alpha)\, \sigma_R^2 + (C_1 \sin
\alpha + C_2 \cos
\alpha)\, \eta^2 \over A_3 \sigma_R^2}}
where as above $A_3$ comes from the dominant terms in
$V_{\Delta\phi}$.

Other interesting possibilities to have spontaneous CP violation are
given in {\branco}, but we have only considered the simplest one.

\listrefs
\end